\documentclass[submitting]{nst}

\usepackage{subfigure,dcolumn}
\usepackage{epstopdf}
\usepackage{mhchem}
\usepackage{upgreek}
\usepackage{hyperref}

\begin{document}

\title{Precise determination of $\ce{^{210}Pb}$ $\beta$ Decay Spectrum at 0 keV and its Implication to Theoretical Calculations\mbox{}}
\thanks{Project supported by the National major scientific research instrument development project (Grant No.11927805)}

\author{Shuo Zhang}
\email[Corresponding author, ]{shuozhang@shanghaitech.edu.cn}
\affiliation{Center for Transformative Science, ShanghaiTech University, Shanghai, 201210, China}

\author{Xavier Mougeot}
\affiliation{CEA, LIST, Laboratoire National Henri Becquerel, CEA-Saclay, Gif-sur-Yvette Cedex, 91191, France}

\author{Hao-Ran Liu}
\affiliation{Division of Ionizing Radiation, National Institute of Metrology, NIM, Beijing, 100029, China}

\author{Ke Han}
\affiliation{INPAC and Department of Physics and Astronomy, Shanghai Jiao Tong University, Shanghai, 200240, China}

\author{Tao Sun}
\affiliation{Shanghai Institute of Microsystem and Information Technology, Chinese Academy of Sciences, Shanghai, 200050, China}

\author{Wen-Tao Wu}
\affiliation{Shanghai Institute of Microsystem and Information Technology, Chinese Academy of Sciences, Shanghai, 200050, China}

\author{Robin Cantor}
\affiliation{STAR Cryoelectronics, 25-A Bisbee Court, Santa Fe, NM, 87508-1338}

\author{Jing-Kai Xia}
\affiliation{Center for Transformative Science, ShanghaiTech University, Shanghai, 201210, China}

\author{Jun-Cheng Liang}
\affiliation{Division of Ionizing Radiation, National Institute of Metrology, NIM, Beijing, 100029, China}

\author{Fu-You Fan}
\affiliation{Division of Ionizing Radiation, National Institute of Metrology, NIM, Beijing, 100029, China}

\author{Bing-Jun Wu}
\affiliation{Shanghai Institute of Microsystem and Information Technology, Chinese Academy of Sciences, Shanghai, 200050, China}

\author{Le Zhang}
\affiliation{School of Physics and Astronomy, Sun Yat-Sen University, Guangzhou, 510297, China}

\author{Ming-Yu Ge}
\affiliation{Institute of High Energy Physics, Chinese Academy of Sciences, Beijing, \quad 100049, China}

\author{XiaoPeng Zhou}
\affiliation{School of Physics, Beihang University, Beijing 100191, China}

\author{Zhi Liu}
\affiliation{Center for Transformative Science, ShanghaiTech University, Shanghai, 201210, China}

\begin{abstract}
Background: The atomic exchange effect will lead to a significant increase in the probability density of $\beta$ decays below a few keV. This effect is very important for scientific experiments such as the solar axion detection that performed by low-energy electron spectroscopy measurements. However, the atomic exchange effect involves multi-electron interactions, especially for a system with 82 electrons such as lead, there are many uncertainties in the model of the atomic exchange effect. Different parameters will lead to different trends in the energy spectrum predicted by the theory, so it is urgent to carry out experimental measurements to provide parameter limits for the theory. Aim: the probability increase brought about by the atomic exchange effect is most obvious near 0 keV, and the $\beta$ energy spectrum is accurately measured near this energy point, so as to provide constraints for the physical model of atomic exchange. Method: However, it is extremely difficult to measure the $\beta$ energy spectrum at 0 keV due to the limitations of electronic noise and internal conversion effects. In order to solve this problem, the excited decay path of $\ce{^{210}Pb}$ was taken as the observation object, by measuring the total energy spectrum of $\beta$ rays and cascaded $\gamma$ rays, the precise measurement of the $\beta$ energy spectrum near 0 keV has been completed. Results: The analysis of the $\beta$ energy spectrum of $\ce{^{210}Pb}$ gives the following conclusions. The experimental results first verified the theory that the exchange effect causes the probability increase at the low energy end near 0 keV. At the same time, the experimental results are higher than the existing predictions of the atomic exchange effect.Conclusion: At least for Pb element, all the electron shells have played a role in improving the probability density of the low end of the $\beta$ energy spectrum, instead of the previous thought that some shells play a role in improving, and some depression. This discovery will promote the theoretical calculation of the $\beta$ energy spectrum of $\ce{^{214}Pb}$, so it has a very positive effect on the background estimation of experiments such as XENON1T, at the same time, it also indicates that the reactor neutrinos have a higher probability density at the omnipotent end. This discovery plays an important role in the development of frontier physics.

\end{abstract}


\maketitle

\section{Introduction}\label{sec.I}

Accurate analysis of $\beta$ energy spectrum is very important for accurate calculation of reaction energy in radionuclide metrology\cite{2007-Broda-Metrologia}, nuclear medicine\cite{2019-Hashempour-JNMB}, dosimetry\cite{1994-Manuel-PMB,1991-Dickens-PMB,1982-Yoshida-JNST} and nuclear industry\cite{2015-Xavier-PRC}, more importantly, it is a key factor limiting the development of frontier fields in physics. The behavior of the $\beta$ energy spectrum at the low energy end has a great influence on dark matter detection experiments\cite{2020-XENON-PRD,2021-Bha-PLB,2020-Ma-JInstP}. The XENON1T experiment is based on electron recoil energy spectroscopy to measure axions from the sun. This experiment is carried out in the deep field large-scale liquid xenon (LXe) facility, and the energy intervals of interest are mainly below a few keV. In 2020, it was reported that a 3$\sigma$ excess phenomenon was measured in the low energy region. Because the interaction cross-section between the solar axion and the detection medium is extremely small, a detection medium with a huge amount of matter is needed to increase the probability of interaction, which inevitably introduces natural radionuclides, among which there are many $\beta$ decay sources. The essence of $\beta$ rays is electrons with continuous energy, which are indistinguishable from the signals of recoil electrons. The $\beta$ decay sources with greater influence include tritium, $\ce{^{214}Pb}$ and $\ce{^{85}Kr}$ nuclides, and the $\beta$ decay types of these nuclides both related with forbidden decay mode. For forbidden decay, the current theoretical and experimental spectra do not align well. Because the electron shell outside the nucleus will introduce the atomic exchange effect during the decay process of $\beta$, the spectral pattern of $\beta$ rays in the low-energy region will change. However, the theory of the atomic exchange effect is not yet perfect, and its behavior at the low-energy end cannot be accurately evaluated, resulting in the inability to verify the accuracy of this exceeding result\cite{2020-XENON-PRD}. Therefore, it is extremely important to study the influence of the outer electron shell on the $\beta$ decay energy spectrum for solar axion detection.

In addition, the study of the behavior of the $\beta$ energy spectrum at the low energy end is crucial as well for reactor neutrino measurement experiments. Upon taking into account the radiation correction, $\beta$ energy spectrum and neutrino spectrum exhibit symmetry, where the latter complements the former. That is, the probability of generation of $\beta$ rays at energy $E$ is equal to the probability of generating neutrinos at $E_0-E$, this means that the precise measurement of the behavior of the $\beta$ spectrum at the low-energy end can give a prediction of the behavior of neutrinos at the high-energy end \cite{2015-Xavier-PRC}. Therefore, for many cutting-edge research in physics, the atomic exchange effect has become a puzzle that must be completed.

Introduction to atomic exchange effect: There is an exchange effect between the decay electron and the extranuclear electron cloud, the electronic wave function of the decay electron overlaps with the bound state wave function, and occupies the position of the bound state electron, causing the electron to be forced to go to the free state of the final state atom, the formula \ref{eq-2} into the form of the formula \ref{eq-3}. This will obviously affect the $\beta$ energy spectrum and decay rate, especially the decay at the low energy end. This effect was proposed by John \cite{1963-John-PR} in 1963, and then Haxton proposed that this effect would cause an enhancement effect at the low energy end, and it was confirmed by subsequent measurements \cite{1985-Haxton-PRL,1992-Harston-PRA}. In 2012, Xavier used $\ce{^{241}Pu}$ to re-verify the effect \cite{2012-Xavier-PRA}. In this work, the author used the MMC microcalorimeter to measure the energy spectrum of $\ce{^{241}Pu}$, and then optimized \cite{2014-Xavier-PRA}. This paper also shows the measurement results of $\ce{^{63}Ni}$, the measurement results show that the exchange effect and shielding effect will lead to a higher probability of the $\beta$ energy spectrum in the low energy region than the traditional theory expects. However, due to the limitation of the trigger threshold of the detector, the deviation phenomenon is only measured in the energy range higher than 0.3keV. Moreover, due to the internal conversion effect and the presence of impurity nuclides, it is impossible to confirm that the deviation is completely caused by the exchange effect. Therefore, it is urgent to find a new measurement mode to accurately measure the energy spectrum near 0keV without the limitation of trigger threshold and other factors, so as to provide guidance for the theoretical verification near this energy region. A further detailed analysis of $\ce{^{45}Ca}$ and $\ce{^{241}Pu}$ is presented in a review paper on computing the $\beta$ spectra of allowed transitions\cite{2018-eendert-rmp}. Recently, Karsten et al. made precise measurements on $\ce{^{151}Sm}$, and found that there is also an excess phenomenon \cite{2022-Karsten-ARI} in the low-energy region. In addition, the author of this article also found that the existing theoretical calculation model, after considering the exchange effect and screening effect, still cannot fit the experimental spectrum well, indicating that the model inside is still not perfect and needs to be improved. Although the exchange effect has been confirmed, its phenomenon of causing excess at the low energy end has also been confirmed. But it is so important for experiments like XENON1T, the unique forbidden transition of $\ce{^{85}Kr}$ and the non-unique first forbidden transition of $\ce{^{212}Pb}$ and $\ce{^{214}Pb}$ Detailed behavior has a significant impact on experimental results. In addition, there is still a mismatch between the experimental and measurement results of the effect itself, so it is necessary to further test the exchange effect. New measurements need to exclude the effects of electronic noise and the decay of excited states of impurity nuclides.

The shape of the $\beta$ energy spectrum can be represented by the formula \ref{eq-1}\cite{1982-Behrens-Book}. It should be emphasized that the natural unit system $\hbar=m_e=c=1$ is used here. Where ${p_e}{E_e}{({E_0}-{E_e})^2}$ is related to the statistical phase space factor, which reflects the momentum distribution relationship between neutrinos and decaying electrons. $E_e$ is the total electron energy, $p_e=\sqrt{{E_e}-1}$.$F(Z,{E_e})$ is called the Fermi function and the shape factor, which represents the electrostatic interaction between decaying electrons and decayed atoms. $C({E_e})$ contains nuclear matrix elements and other factors related to lepton energy. For the allowed transition $C({E_e})$ is energy-independent, and Xavier mentioned that when $2\xi/{E_0}>100$ is satisfied, the forbidden transition can be approximately treated as an allowed transition \cite{2015-Xavier-PRC}. At this point, the formula \ref{eq-1} can be expressed as the formula \ref{eq-2}.

\begin{equation}\label{eq-1}
  \frac{dN}{dE_e}\propto{p_e}{E_e}{({E_0}-{E_e})^2}F(Z,{E_e})C({E_e})
\end{equation}

\begin{equation}\label{eq-2}
  \frac{dN}{dE_e}\propto{p_e}{E_e}{({E_0}-{E_e})^2}F(Z,{E_e})
\end{equation}

\begin{equation}\label{eq-3}
  \frac{dN}{dE_e}\propto{p_e}{E_e}{({E_0}-{E_e})^2}F(Z,{E_e})*[1+{{\eta}^T(E_e)}]
\end{equation}

The atomic exchange effect predicts that the atomic shell has an impact on the probability of $\beta$ decay, and different shells contribute different probabilities. According to the current mainstream theory, the contribution of the some shell to the $\beta$ decay is positive, and the contribution of the other shell to the decay is negative. However, according to Nitescu's point of view, the wave function of decaying electrons must be orthogonalized with the energy level after decay, and the calculation results after orthogonalization show that all the core shells play a positive role in $\beta$ decay. Moreover, the probability improvement effect will increase by nearly 20\% or even higher near 0keV \cite{2023-Nitescu-PRC}, that is, the atomic exchange effect has a very obvious effect at 0keV, and the best verification point is also at 0keV.  %(Xavier, please give a brief introduction to the reason why $\ce{^{210}Pb}$ all nuclear shells can enhance the $\beta$-decay. Please add a introduction to Po.+Ne. model, which shell is positive which shell is negative, and if it is convenient, please introduce the selection rule for which one is positive and which one is negative)

\begin{figure}[!htb]
\includegraphics[width=0.9\hsize]{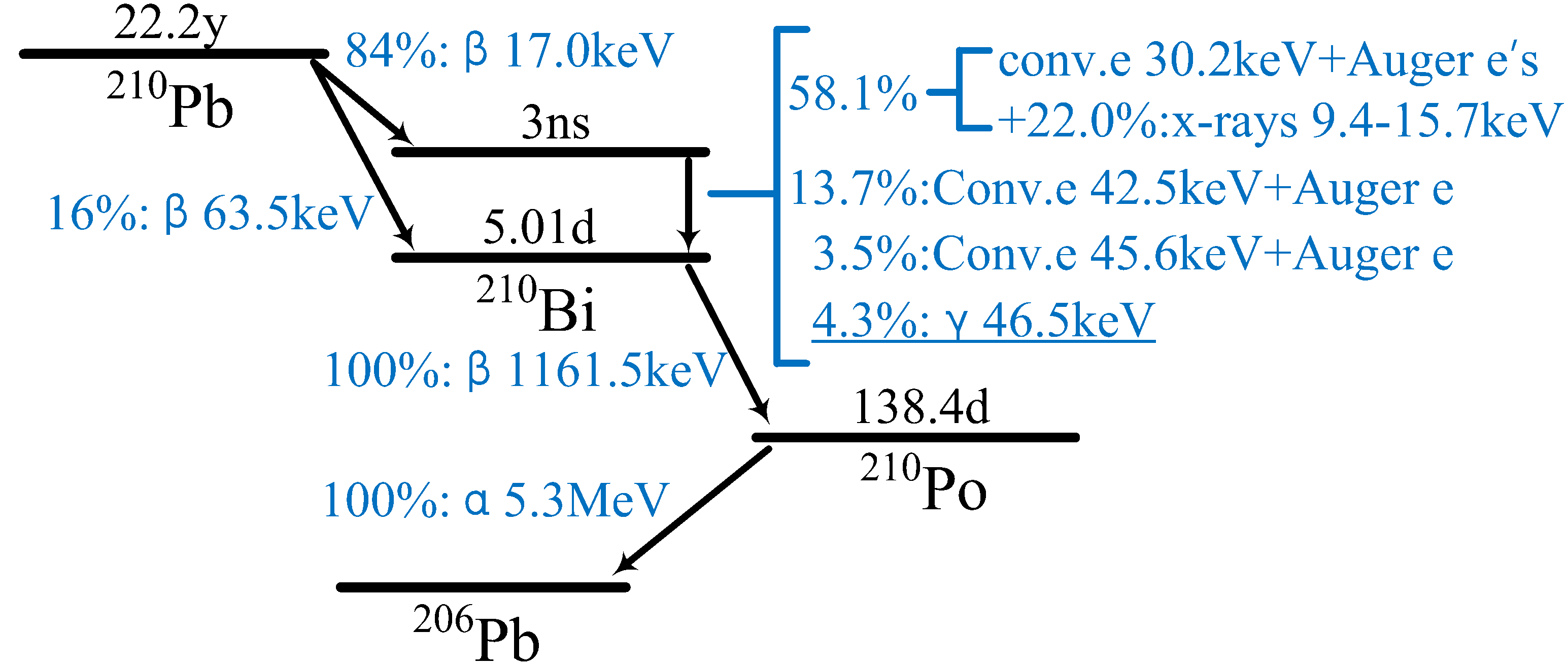}
\caption{decay scheme.}
\label{fig0}
\end{figure}

The decay of $\ce{^{210}Pb}$ to $\ce{^{210}Bi}$ is a non-unique forbidden transition, it has two decay paths and two $\beta$ ray decay paths, The total decay energy is 63.5keV. A decay channel directly decays to the ground state of $\ce{^{210}Bi}$, and the highest energy spectrum of the emitted $\beta$-ray is 63.5keV\cite{2020-Alekseev-PRC}. The energy spectrum of the $\beta$-ray emitted by the excited decay channel has the highest energy of 17keV, decays to the excited state of $\ce{^{210}Bi}$, and after about 3ns de-excites, it releases a single-energy $\gamma$ photon of 46.5keV . In this study, the branch with cascaded $\gamma$ rays was selected, and the TES-based microcalorimeter was used to measure the cascade spectrum, which eliminated the influence of impurity nuclide excitation decay channels to the greatest extent. At the same time, the problem of the trigger threshold is avoided, so the $\beta$ energy spectrum of 0keV is accurately measured. The effect of cascading decay on this measurement will be explained and analyzed in detail in the section on systematic bias. After comparison, the experimental measurement results verify that in the extremely low energy region, the exchange effect will lead to an increase in the probability of the $\beta$ energy spectrum, which verifies the previous theoretical and experimental measurement expectations, and extends the expectation to 0keV. Experimental results show that all atomic shells, at least for the element Pb, contribute to a boost to the probability density at the lower end of the beta spectrum. This phenomenon is also applicable to $\ce{^{214}Pb}$.

\section{Measurement and Analysis}\label{sec.II}

\subsection{Introduction to Measuring Devices}\label{A}

\begin{figure}[!htb]
\includegraphics[width=0.9\hsize]{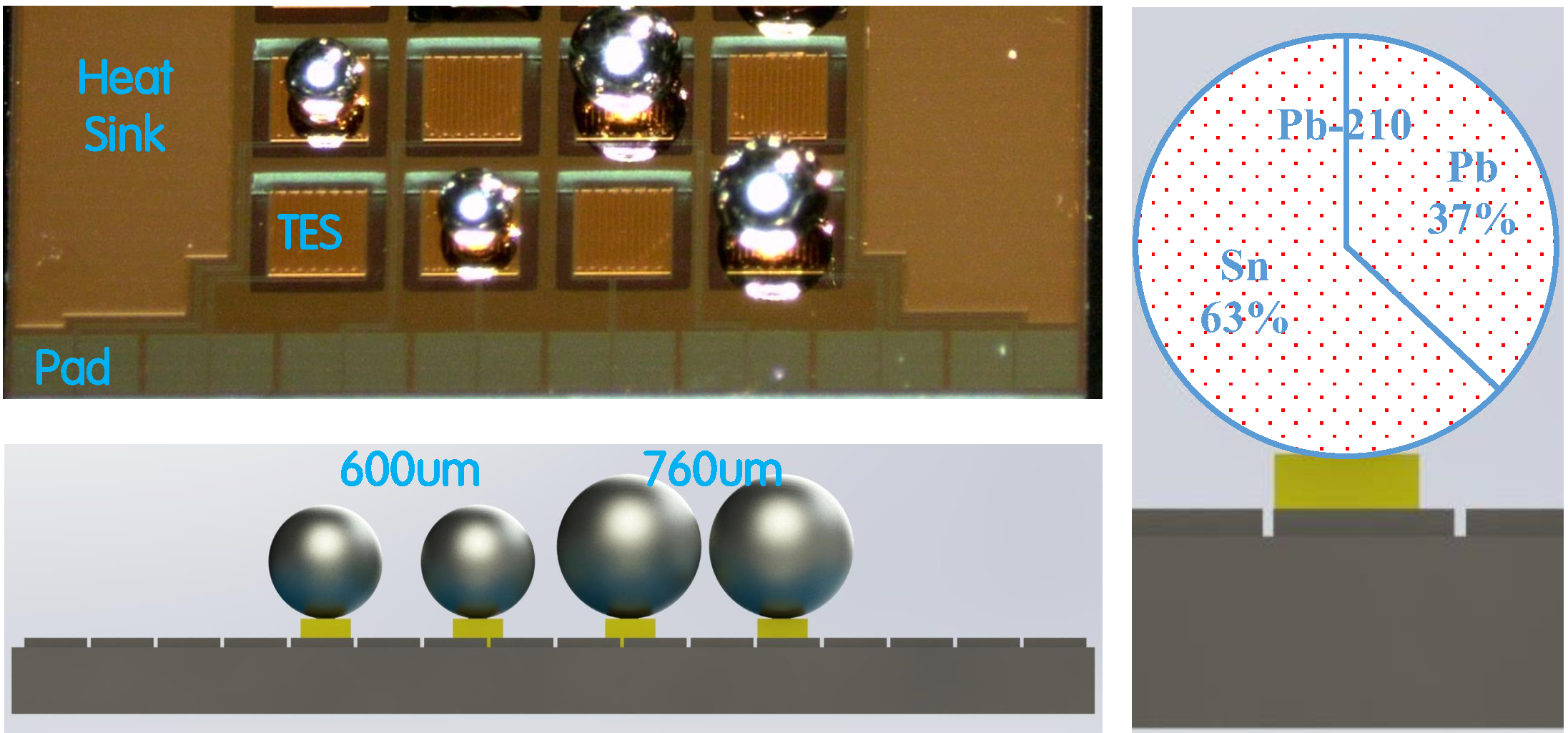}
\caption{TES detector core structure.}
\label{fig1a}
\end{figure}

The core structure of the TES detector is shown in figure \ref{fig1a}. On one side of a 16-pixel TES chip, four lead-tin alloy balls with a diameter of 600 microns were glued, and four lead-tin alloy balls with a diameter of 760 microns were glued on the other side. The adhesive is Stycast 2850, the glue dot size is about 100 microns, and the thickness is about 10 microns. For brevity, only half of the chip on the wire bonding side is shown here. The core of the TES chip is a TES film composed of a molybdenum-copper double-layer film. The superconducting transition edge temperature is about 70mK, the side length of the film is 500 microns, and the wire material is molybdenum. The lead-tin alloy ball is composed of 37\% lead and 63\% tin. According to the Bloch formula calculation and Geant4 simulation calculation, about 99.99\% of the 17keV electrons in the 600um diameter lead-tin alloy ball will be absorbed. Because the entire energy spectrum will be corrected by this curve when calculating the energy spectrum in the later stage.

\begin{figure*}[!htb]
\includegraphics[width=0.9\hsize]{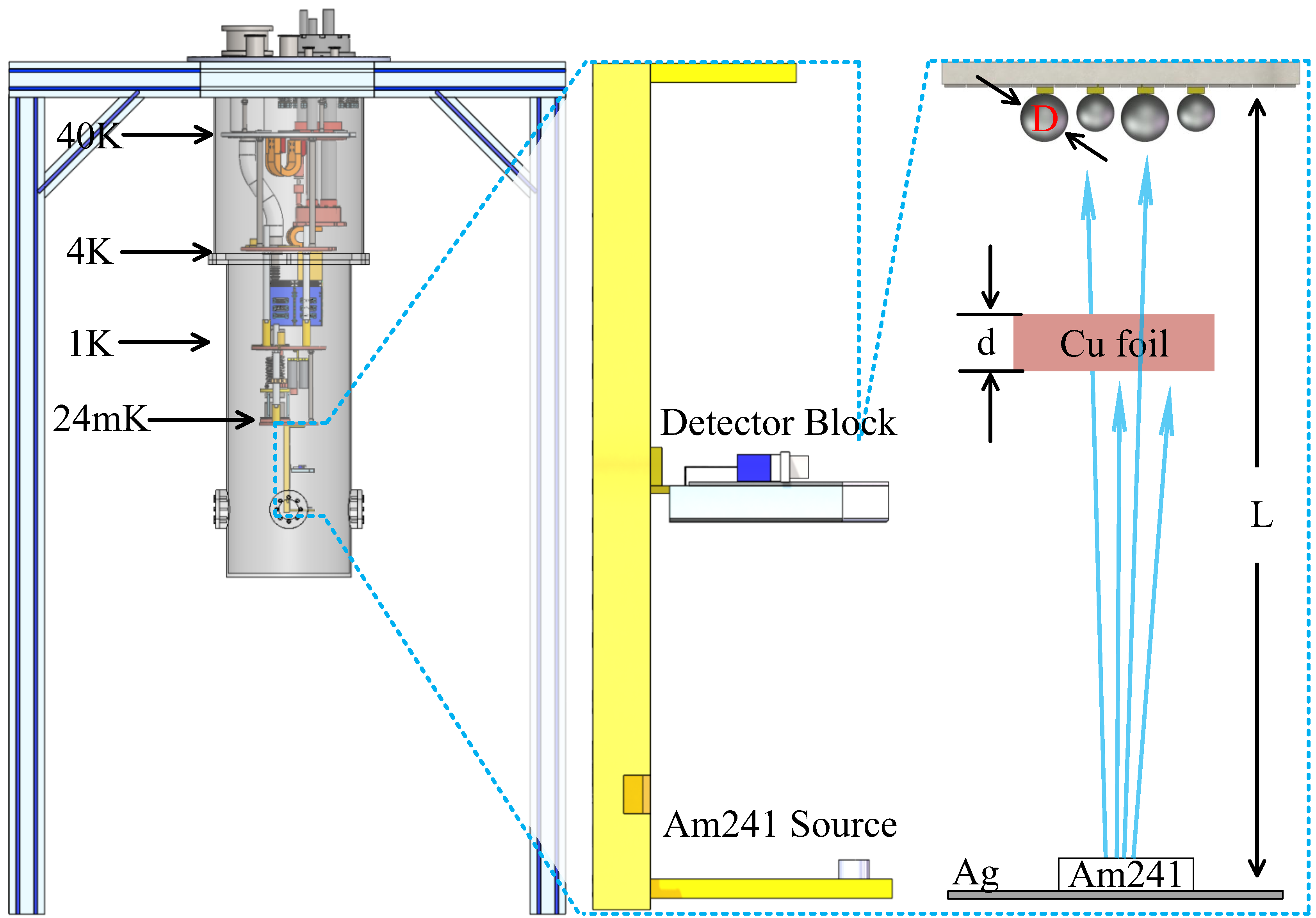}
\caption{Schematic diagram of measurement system.}
\label{fig1}
\end{figure*}

As shown in \ref{fig1}, the measurement is carried out in a dilution refrigerator, and the TES detector is covered by a light-shielding structure to avoid spurious signals introduced by scintillation photons. Both the TES detector and the $\ce{^{241}Am}$ source used for calibration are installed on a heat-conducting copper plate, and a copper plate is installed between them. By changing the thickness, the brightness of the calibration source can be flexibly adjusted.

\subsection{Systematic Deviation Analysis}\label{B}

The systematic deviation mainly comes from the absorber, and $\ce{^{210}Pb}$ is evenly distributed throughout the absorber. Three main factors need to be considered, firstly, the absorber cannot fully absorb the $\beta$ decay energy, that is, the energy escape will bring deformation to the $\beta$ energy spectrum, and secondly, the background energy spectrum caused by the cascading decay. Finally, $\gamma$ rays with energy of 59.54keV from $\ce{^{241}Am}$, Compton scattering and X-ray escape will form characteristic peaks and continuum near the all-energy peak, which will be superimposed on the $\beta$ energy spectrum.

As is mentioned above, this work selects the excited decay channel of $\ce{^{210}Pb}$ as the measurement object, and the decay process is shown in detail in figure\ref{fig0}.The nuclear decay process of $\ce{^{210}Pb}$ to $\ce{^{210}Bi}$ produces neutrinos, $\beta$ rays, and then $\ce{^{210}Bi}$ transition to the ground state.This process produces $\gamma$ rays with an energy of 46.5keV with a probability of 4.3\%, and also produces internal conversion electrons, Auger electrons and X-rays with a total energy of 46.5keV with a certain probability. The penetrating power of internal conversion electrons, Auger electrons and X-rays (15.7keV) is extremely weak, the penetration depth is on the order of tens of nanometers, and the diameter of the absorber reaches 760 microns, so all their energy will be completely absorbed. For the $\gamma$ ray of 46.5keV, its penetrating power is stronger. After calculation, the escape probability of this $\gamma$ ray in the lead-tin alloy with a diameter of 760 microns is less than 10\%. Combined with its emission probability of 4.3\%, the escape probability of this part of energy is about 0.4\%. For the convenience of discussion, the particles produced in the de-excitation process are referred to as de-excitation secondary particles. The probability of neutrinos interacting with the absorber is extremely low, and it can be considered that all neutrinos escape. The highest energy of $\beta$ rays is only 17keV, and the penetration is lower than 10nm, which can be considered as full absorption. Overall, $\beta$ rays and de-excited secondary particles are totally absorbed to form a continuous energy spectrum starting at 46.5keV, with an energy spectrum width of 17keV, and the escape of 0.4\% related to 46.5keV$\gamma$ rays , the 0.4\% excited decay count will be superimposed on the count of the ground state decay track. Since the spectral width of the excited decay track is only 17keV, which is much lower than 46.5keV, it will not affect the shape of the target observed spectrum.

The effect of cascade decay: $\ce{^{210}Pb}$ decays into $\ce{^{210}Bi}$ through $\beta$, $\ce{^{210}Bi}$ is also a $\beta$ radioactive source, with a half-life of 5 days, which is much lower than the half-life of $\ce{^{210}Pb}$, and the lifetime of the lead-tin alloy used in this experiment It has been far more than 5 days, so a decay equilibrium will be formed, that is, the decay rates of $\ce{^{210}Pb}$ and $\ce{^{210}Bi}$ are equal.The decay energy of $\ce{^{210}Bi}$ is 1161keV, much larger than the decay energy of $\ce{^{210}Pb}$, the average probability density is about 0.054 of $\ce{^{210}Pb}$.At the same time, the decay probability of $\ce{^{210}Bi}$ changes smoothly with energy, and a background close to a slanted line will be formed between 0-63.5keV.$\ce{^{210}Bi}$ decays into $\ce{^{210}Po}$, $\ce{^{210}Po}$ becomes $\ce{^{206}Pb}$ in the form of $\alpha$ decay, with a half-life of 138 days, similar to $\ce{^{210}Bi}$ , $\ce{^{210}Po}$ will form a decay equilibrium with $\ce{^{210}Bi}$ and $\ce{^{210}Pb}$, and the decay rates of the three are close. Since $\ce{^{210}Po}$ undergoes $\alpha$ decay, all the decay energy 5407.53keV will be completely absorbed by the absorber, producing a single-energy signal and a line on the energy spectrum, because 5407.53keV is much larger than the saturation energy of this detector, so many saturation signals will be seen in this measurement, and these signals will not have any impact on this research.

The influence of the calibration source: this work will select $\ce{^{241}Am}$ as the calibration source, and use aluminum foil and copper foil to filter out the $\alpha$ particles generated by it, and mainly use its energy of 59.54keV $\gamma$ rays, Compton scattering and X-ray escape will form characteristic peaks and continuum near the full energy peak, thus superimposed on the $\beta$ energy spectrum. The influence of the escape characteristic peak needs to evaluate the escape probability of X-rays with different energies by measuring the emission line of the absorber material, so as to obtain the relative intensity of the escape characteristic peak. The influence of Compton scattering on the energy spectrum comes from two parts. The first part is that the $\gamma$ ray hits the absorber, and the secondary photon escapes to bring energy loss. Since the photon at 46.5keV mainly occurs the photoelectric effect, the Compton scattering cross section is 4 orders of magnitude lower than the photoelectric effect, so the influence of the energy spectrum deformation brought by the first part is minimal. Another situation is that $\gamma$ rays hit the metal and other materials around the absorber, and the secondary photons after Compton scattering hit the absorber, thus forming a continuous energy spectrum with a specific shape. This article has given the shape of the energy spectrum through geant4 simulation, but there is a certain deviation from the measured spectrum shape. In the actual measurement, the influence of this part is reduced to less than 1\% by extending the distance, and the Compton effect is obvious. Part of the data is discarded, which has little impact on the theoretical verification of the atomic exchange effect, and the simulation results of geant4 are not given here.

\subsection{Experimental Measurements and Results}\label{C}

\begin{figure*}[!htb]
\includegraphics[width=0.9\hsize]{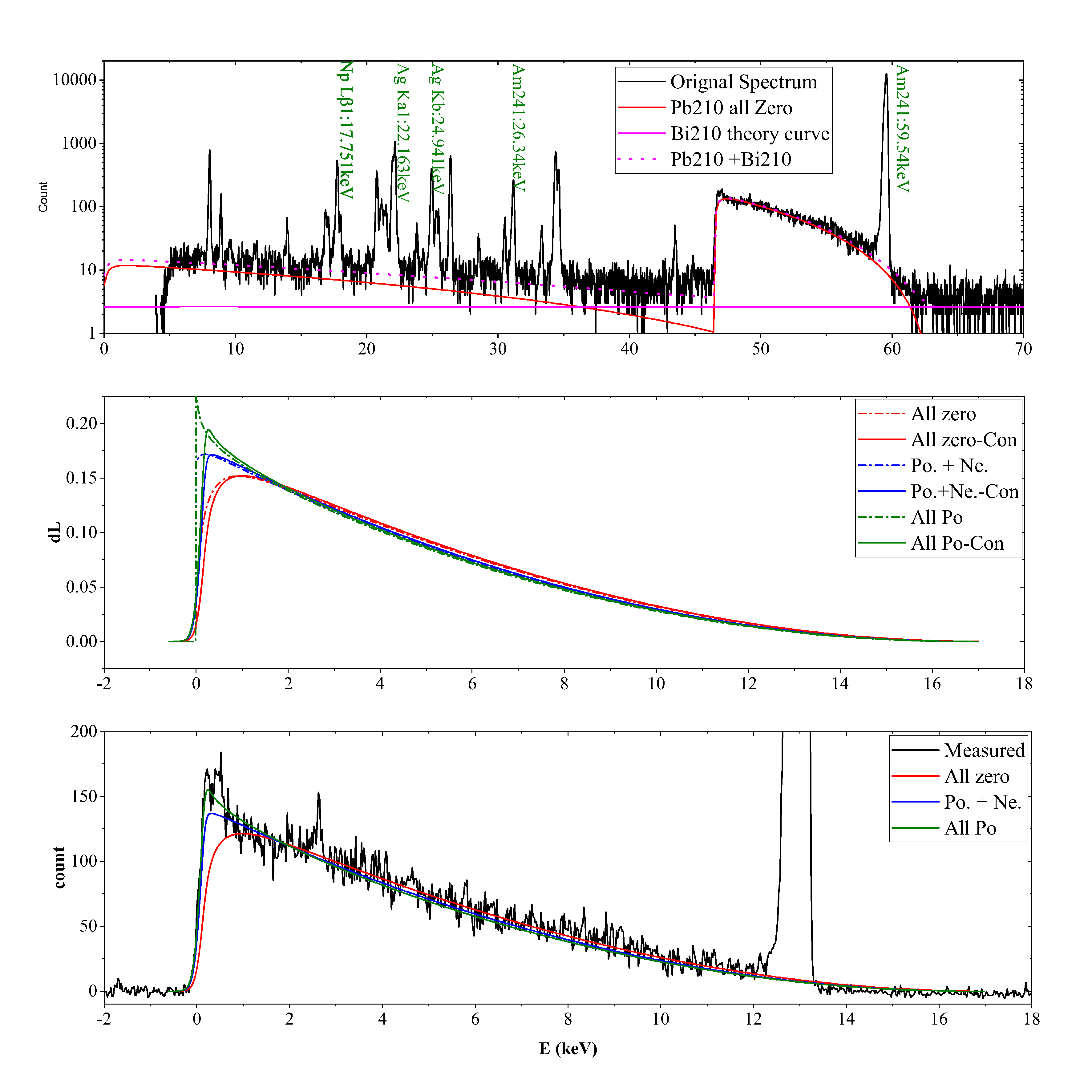}
\caption{Energy spectrum of 760 micron lead-tin alloy ball.}
\label{fig2}
\end{figure*}

Place a $\ce{^{241}Am}$ source about 10-22cm away from the TES, cover it with 2 layers of copper foil and a layer of aluminum foil to block most $\alpha$ particles. Let only $\gamma$ rays and X-rays shine on the pewter ball of TES. After about 30 days of measurement, the energy spectrum as shown in \ref{fig2} was obtained.On the energy spectrum at 59.54keV and 26.34keV, you can clearly see the $\gamma$ spectral line of $\ce{^{241}Am}$, and the Np after $\ce{^{241}Am}$ decay L line. In addition, a peak originating from the silver element can be seen, which mainly originates from the substrate being excited by the radioactive source. In addition, it can be seen that starting from 46.5keV, there is a continuous energy spectrum, which is the energy spectrum produced by the cascade of 46.5keV $\gamma$ rays and 17keV $\beta$ rays. Firstly, according to the trend of the first 63.5keV energy $\beta$ spectrum, its background to the 17keV energy spectrum is subtracted. Then, according to the background height between 40keV and 70keV, it is assumed that it is a linear background and removed. After background subtraction, subtract 46.5keV from the abscissa to get the $\beta$ energy spectrum of the 17keV branch.

\begin{figure*}[!htb]
\includegraphics[width=0.9\hsize]{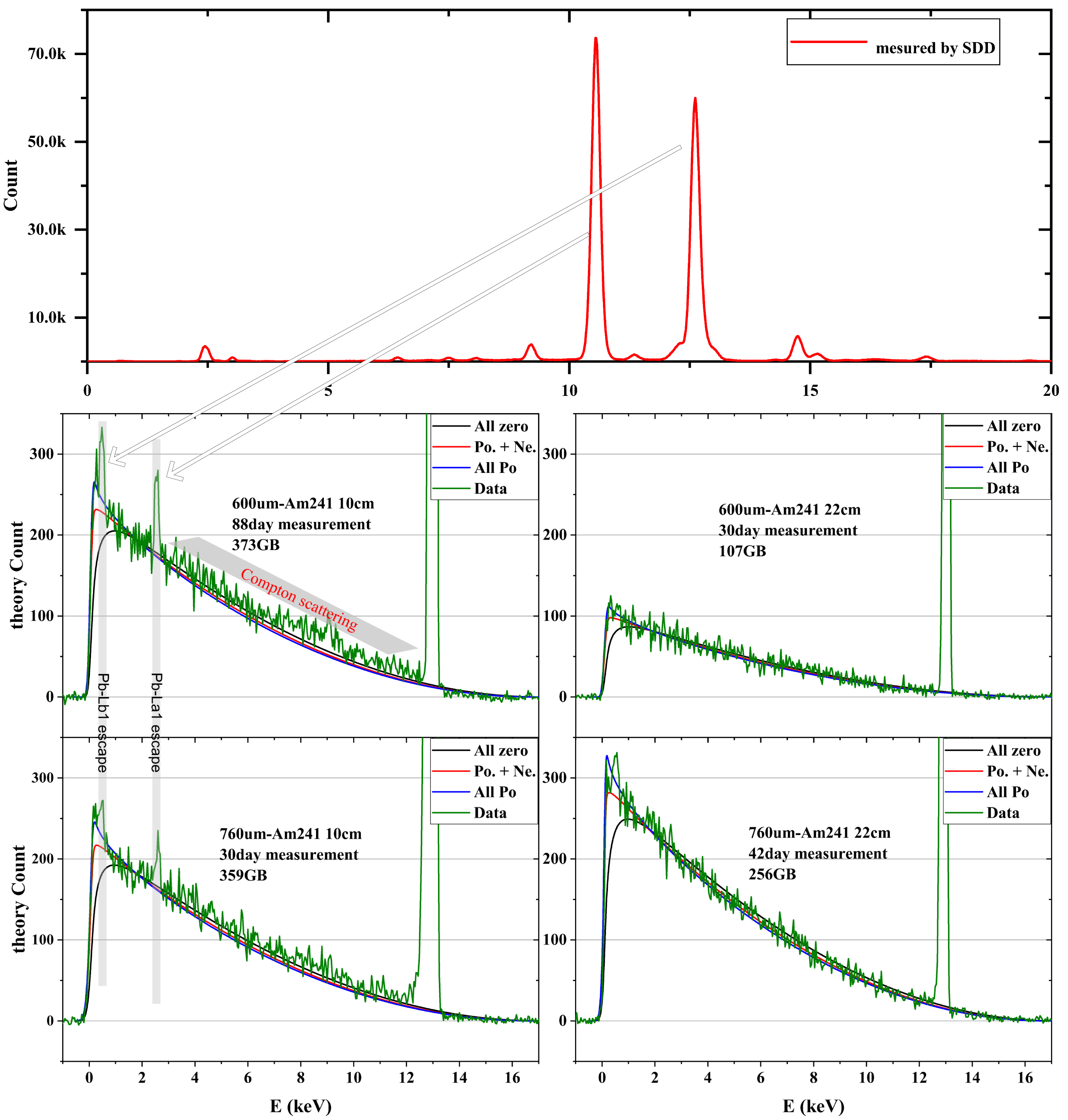}
\caption{Comparison of theoretical energy spectrum of measured results.}
\label{fig3}
\end{figure*}

Firstly, at a distance of about 10cm from $\ce{^{241}Am}$, the energy spectrum collection work is performed on the 600 micron and 760 micron lead-tin alloy pellets.According to the figure \ref{fig3}, it can be seen that there are two obvious peak structures at 0.427keV and 2.489keV. Lead element absorbs $\gamma$ rays at 59.54keV to produce characteristic X-rays, which escapes from the lead-tin alloy absorber and leads to this result.The escaped characteristic lines mainly include $\ce{L_{{\beta}1}}$ (corresponding to 0.427keV, that is, 59.54keV minus 12.614keV and then minus 46.5keV) and $\ce{L_{{\alpha }1}}$ (corresponding to 2.489keV, that is, 59.54keV minus 10.551keV and then minus 46.5keV). In order to verify whether lead has other obvious escape characteristic peaks, X-ray fluorescence measurement was done on lead element. It can be seen that $\ce{L_{{\alpha}1}}$ and $\ce{L_{{\beta}1}}$ are the two brightest spectral lines, and the other spectral lines are relatively faint, with a height of only 10\% of the two main emission lines, so in general, only need to pay attention to these two escape characteristic peaks on the continuum.In fact, on the top energy spectrum of figure \ref{fig2}, $\ce{K_{{\alpha}1}}$ and $\ce{K_{{\ alpha}2}}$ escape characteristic peak. Since these two escape peaks correspond to 25.27keV and 25.04keV, which are much higher than the $\ce{L_{{\alpha}1}}$ energy of lead, and also have a stronger escape ability, so their intensity is much greater than that of lead two escape peaks.However, these two spectral lines do not appear on the $\beta$ spectrum of $\ce{^{210}Pb}$, and have no influence on this study. In addition, it can be seen in the figure \ref{fig3} that at $\gamma$ close to the energy of 59.54keV, there is a part of the continuous spectrum beyond the model, and the count contribution is lower than 30count per channel.This part of the energy spectrum is mainly due to Compton scattering. Although the counts in this part are not as obvious as the escape characteristic peaks of lead, they are continuous spectra and cannot be easily deducted. The only way to do this is by pulling the radioactive source farther away, until the contribution of Compton scattering to the measurement is negligible.

Therefore, when the radioactive source is pulled away to 22cm, it is found that the escape characteristic peak basically disappears, and the continuum background brought by Compton scattering also basically disappears. In order to obtain a better fitting effect, a TES device with a diameter of 760 microns was re-made, which has better energy resolution. In order to reduce the contribution of the bottom, a thicker copper foil was replaced between the two, and the flux of the calibration source was reduced by about 4 times, thereby basically eliminating the influence of the escape characteristic peak and the Compton scattering background.

Three theoretical curves are taken respectively, that all electron shells have no influence on $\beta$ decay (All zero), some electron shells have positive influence on decay (Po.+Ne.), and all electron shells have positive influence on decay (All Po.). After deducting the background using the energy spectrum of $\ce{^{241}Am}$ at 59.54keV, it is convolved with three theoretical curves to obtain three convolved theoretical curves. The preliminary analysis of the energy spectrum at 59.54keV has an energy resolution of about 180eV, but there is a certain asymmetry. Therefore, the Gaussian function cannot be used to directly convolve the three theoretical curves when performing convolution. Directly use the contour of 59.54keV to convolve the three theoretical curves, and then perform normalization. Compare the theoretical curve after convolution with the measured results. In order to remove the influence of Compton scattering, remove the data between 4.5-13.5keV, do the fitting, and calculate the chi-square. The calculation results are shown in the table\ ref{tb1}, it is found that the curve fitting degree of All Po. is the best.

\begin{figure*}[!htb]
\includegraphics[width=0.9\hsize]{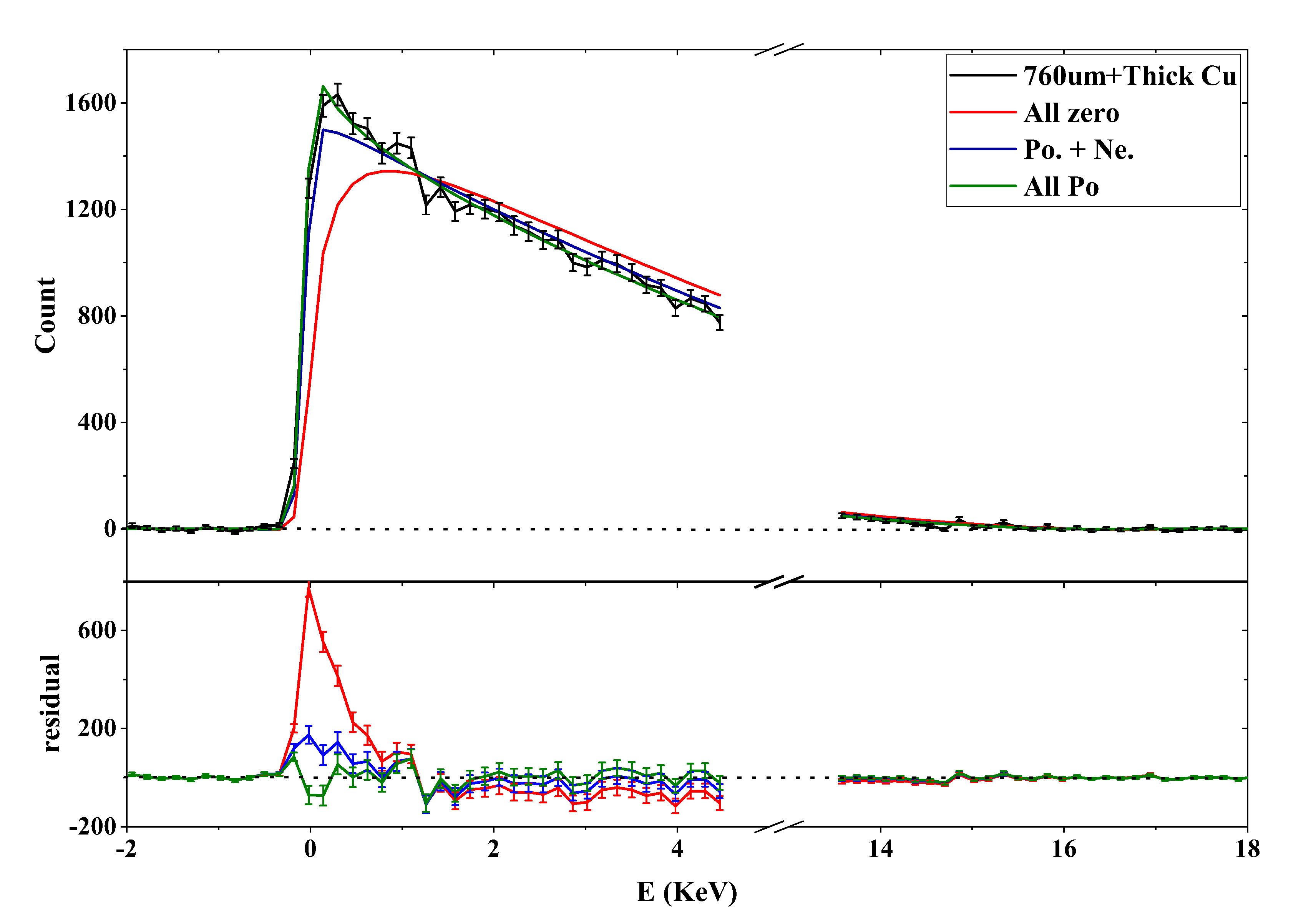}
\caption{Comparison curves of measured energy spectrum and three physical models.}
\label{fig4}
\end{figure*}

\begin{table}
\begin{tabular}{|c|c|c|c|c|}
\hline
\textbf{Sta.result\textbackslash{}Model} & \textbf{All zero} & \textbf{Po.+Ne.} & \textbf{All Po.} & \textbf{Notes}    \\ \hline
\textbf{chi square}                      & 1197.0            & 255.5            & 180.3            & data point is 132 \\ \hline
\textbf{reduced chi square}              & 9.068             & 1.936            & 1.366            & data point is 132 \\ \hline
\end{tabular}
\label{tb1}
\caption{Statistical results of measured energy spectrum and three physical models.}
\end{table}

\section{Summary and Outlook}\label{sec.IV}
All the outer atomic shells, at least for Pb elements, have played a role in raising the probability density at the lower end of the $\beta$ energy spectrum. Contrary to the previous assumption that specific sections of the shell have either an elevating or suppressing effect. These findings have direct implications in guiding the theoretical calculation of the $\beta$ energy spectrum of $\ce{^{214}Pb}$, thereby significantly impacting the background estimation of experiments such as XENON1T. Simultaneously, it also indicates an increased probability density of reactor neutrinos towards the higher energy range. These results have significant implications for advancing frontier physics research.

\section{Thanks}\label{sec.V}

Many thanks to Dan Maccammon of University of Wisconsin-Madison, He Xiaotao and He Qinghua of Nanjing University of Aeronautics and Astronautics, Yao Jiangming of Sun Yat-sen University, Dai Xiongxin of China Radiation Institute, Zhang Guoqiang of Institute for Advanced Study, Ma Long of Fudan University, Yang Xiaofei of Peking University, and beyer of PTB and other professionals for detailed discussions.

Supported by the National Natural Science Foundation of China\textbf{National Major Scientific Research Instrument Development Project} (approval number 11927805).

\end{document}